\documentclass[a4paper]{article}

\usepackage{INTERSPEECH_v2}
\usepackage{multirow} 
\usepackage{color}

\title{Conditional Generative Adversarial Networks for Speech Enhancement and Noise-Robust Speaker Verification}
\name{Daniel Michelsanti and Zheng-Hua Tan}
\address{
  Department of Electronic Systems, Aalborg University, Denmark}
\email{dmiche15@student.aau.dk, zt@es.aau.dk}

\begin{document}

\maketitle
\thispagestyle{myheadings}
\markright{INTERSPEECH 2017\hfill August 20-24, 2017, Stockholm, Sweden\hfill}
\pagestyle{myheadings}
\markright{INTERSPEECH 2017\hfill August 20-24, 2017, Stockholm, Sweden\hfill}
\begin{abstract}

\noindent Improving speech system performance in noisy environments remains a challenging task, and speech enhancement (SE) is one of the effective techniques to solve the problem. Motivated by the promising results of generative adversarial networks (GANs) in a variety of image processing tasks, we explore the potential of conditional GANs (cGANs) for SE, and in particular, we make use of the image processing framework proposed by Isola et al. \cite{isola2016image} to learn a mapping from the spectrogram of noisy speech to an enhanced counterpart. The SE cGAN consists of two networks, trained in an adversarial manner: a generator that tries to enhance the input noisy spectrogram, and a discriminator that tries to distinguish between enhanced spectrograms provided by the generator and clean ones from the database using the noisy spectrogram as a condition. We evaluate the performance of the cGAN method in terms of perceptual evaluation of speech quality (PESQ), short-time objective intelligibility (STOI), and equal error rate (EER) of speaker verification (an example application). Experimental results show that the cGAN method overall outperforms the classical short-time spectral amplitude minimum mean square error (STSA-MMSE) SE algorithm, and is comparable to a deep neural network-based SE approach (DNN-SE).

\end{abstract}

\section{Introduction}

\noindent Dealing with degraded speech signals is a challenging yet important task in many applications, e.g. automatic speaker verification (ASV) \cite{kolboek2016speech}, speech recognition \cite{seltzer2013investigation}, mobile communications and hearing assistive devices \cite{chen2016large, lu2013speech, kolbaek2017speech}. When the receiver is a human user, the objective of SE is to improve quality and intelligibility of noisy speech signals. When it is an automatic speech system, the goal is to improve the noise-robustness of the system, e.g. to reduce the EERs of an ASV system under adverse conditions. 
In the past, this problem has been tackled with statistical methods like Wiener filter and STSA-MMSE \cite{loizou2013speech}. Lately, deep learning methods have been used, such as DNNs \cite{kolbaek2017speech, xu2015regression}, deep autoencoders (DAEs) \cite{lu2013speech}, and convolutional neural networks (CNNs) \cite{park2016fully}. However, to our knowledge, no one has tried to use GANs for SE yet.

GANs are a framework recently introduced by Goodfellow et al. \cite{goodfellow2014generative}, which consists of a generative model, or generator (G), and a discriminative model, or discriminator (D), that play a min-max game between each other. In particular, G tries to fool D which is trained to distinguish the output of G from the real data. 
The architectures used in most of the works today \cite{goodfellow2016nips} are based on  deep convolutional GAN (DCGAN)  \cite{radford2015unsupervised} that successfully tackles training instability  issues when GANs are applied to high resolution  images. Three key ideas are used to accomplish this goal. First, batch normalization \cite{ioffe2015batch} is applied to most of the layers. Then, the networks are designed to have no pooling layers as done in~ \cite{springenberg2014striving}. Finally, the training is performed adopting the Adam optimizer~\cite{kingma2014adam}.

So far GANs have been successfully applied to a variety of computer vision and image processing tasks \cite{isola2016image, radford2015unsupervised, ledig2016photo, zhang2017image}. However, their adoption for speech-related tasks is rare with one exception in \cite{mobin2016voice}, in which the authors of the report applied a deep visual analogy network \cite{reed2015deep} as a generator of a GAN for voice conversion, and the results are presented as example audio files without speech quality or intelligibility or other measures.
In a related field, for music, the GAN concept was applied to train a recurrent neural network for classical music generation~\cite{mogren2016c}.


Very recently, a general-purpose cGAN framework called Pix2Pix was proposed for image-to-image translation \cite{isola2016image}. Motivated by its successful deployment on several tasks, we adapt the framework in this work, aiming to explore the potential of cGANs for SE, as part of the overall goal of investigating the feasibility and performance of GANs for speech processing. Specifically, we use Pix2Pix to learn a mapping between noisy and clean speech spectrograms as well as to learn a loss function for training the mapping.



\section{Pix2Pix framework for speech enhancement}
\noindent In GANs, G represents a mapping function from a random noise vector $\mathbf{z}$ to an output sample $G(\mathbf{z})$, ideally indistinguishable from the real data $\mathbf{x}$ \cite{goodfellow2014generative}. 
%
%
In cGANs, both G and D are conditioned on some extra information  $\mathbf{y}$ \cite{isola2016image}, and they are trained following a min-max game with the objective:
\begin{equation}
\begin{split}
	L&(D, G) =  \mathbb{E}_{\mathbf{x},\mathbf{y} \sim\ p_{data}(\mathbf{x},\mathbf{y})} [\log(D(\mathbf{x}, \mathbf{y}))] +\\				    &\mathbb{E}_{\mathbf{z} \sim\ p_\mathbf{z}(\mathbf{z}), \mathbf{y} \sim\ p_{data}(\mathbf{y})} [\log(1 - D(G(\mathbf{z}, \mathbf{y}),\mathbf{y}))].
\end{split}
\end{equation}

\begin{figure}[ht]
  \centering
  \includegraphics[width=\linewidth]{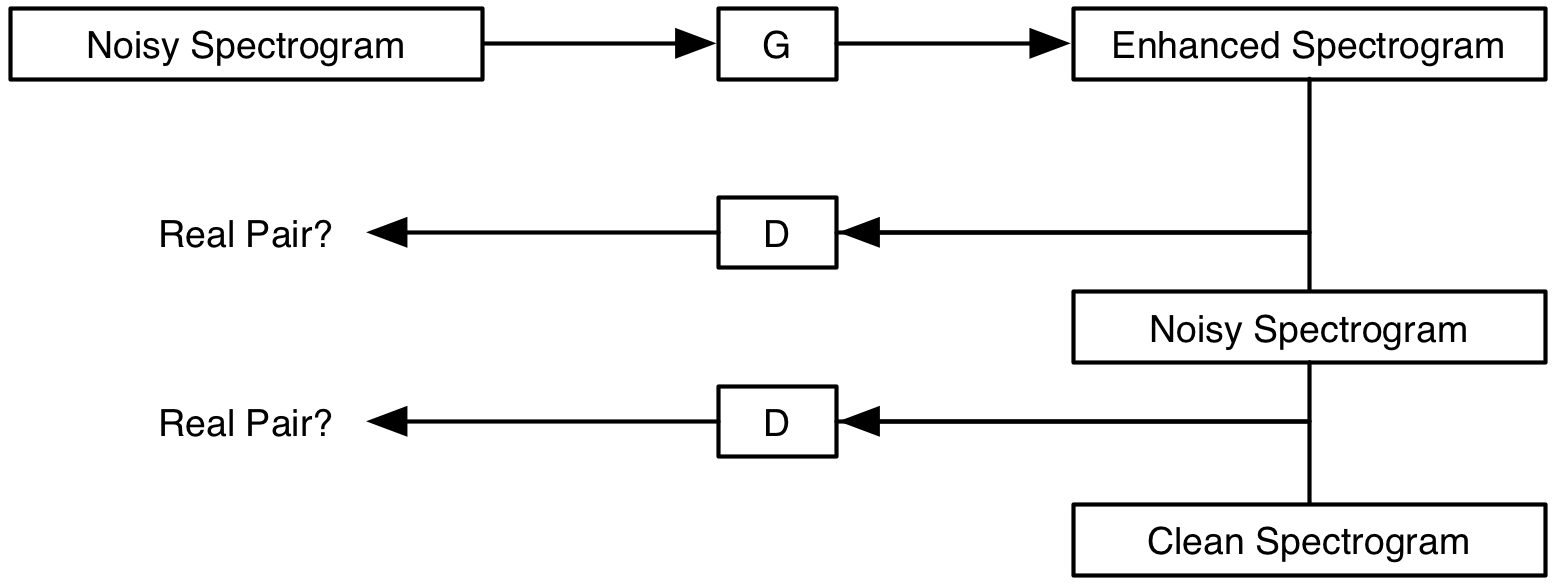}
  \caption{Generator (G) and discriminator (D) in the Pix2Pix framework for speech enhancement. G generates an enhanced spectrogram from a noisy input by fooling D, which tries to classify a spectrogram as clean or enhanced, conditioned on the respective noisy spectrogram.}
  \label{fig:pix2pix}
\end{figure}

Pix2Pix differs from other cGAN works, like \cite{mirza2014conditional}, because it does not use $\mathbf{z}$. Isola et al. \cite{isola2016image} report that adding a Gaussian noise as an input to G, as done in \cite{wang2016generative}, was not effective. Hence, they introduce noise in the form of dropout, but this technique failed to produce stochastic output. However, we are more interested in an accurate mapping between a noisy spectrogram and a clean one than a cGAN able to capture the full entropy of the distribution it models, so this represents a minor issue. Figure \ref{fig:pix2pix} shows how the data and the condition are used during training in the particular case of this paper.

In addition to the adversarial loss $L(D, G)$ that is learned from the data, Pix2Pix utilizes also L1 distance between the output of G and the ground truth. The choice of combining different losses, like L2 distance \cite{pathak2016context} or perceptual losses  for a specific task \cite{ledig2016photo, zhang2017image}, has been shown to be beneficial. In Pix2Pix, L1 distance is preferred to L2 because it encourages less blurring \cite{isola2016image} and it tends to generalize better if compared to perceptual losses.

Furthermore, G and D, adapted from \cite{radford2015unsupervised}, are a U-Net \cite{ronneberger2015u} and a PatchGAN, respectively. Since in image-to-image translation tasks, the input and the output of G share the same structure,  G is an encoder-decoder where each feature map of the decoder layers is concatenated with its mirrored counterpart from the encoder to avoid that the innermost layer represents a bottleneck for the information flow. Besides, D is built to model the high frequencies of the data, as the low frequency structure is captured by the L1 loss. This is achieved by considering local image patches. In particular, D is applied convolutionally across the image to classify each patch as real or fake. Then, the obtained scores are averaged together to get a single output. This architecture has the advantage of being smaller and can be applied on images of different sizes \cite{isola2016image}. When the patch size of D has the same size of the input image, D is equivalent to a classical GAN discriminator.

Our Pix2Pix implementation is based on \cite{lin2016pix2pix}, with G that gets a $256 \times 256$ 1-channel image, while D a $256 \times 256$ 2-channel image. The main differences with the original framework are the adoption of $5 \times 5$ filters in the convolutional layers, and the last layer of D which is flattened and fed into a single sigmoid output as in \cite{radford2015unsupervised}.



\subsection{Preprocessing and training}
\noindent For speech signals with a sample rate of 16 kHz, we computed a time-frequency (T-F) representation using a 512-point short time Fourier transform (STFT) with a hamming window size of 32 ms and a hop size of 16 ms. In this way, the frequency resolution is 16 kHz / 512 = 31.25 Hz per frequency bin. We considered only the 257-point STFT magnitude vectors which cover the positive frequencies due to symmetry. Our generator G accepts $256\times256\times1$ input, so for training we concatenated all the speech signals and then split the spectrogram every 256 frames, while for testing we zero-padded the spectrogram of each test sample in order to have the number of frames equal to a multiple of 256 and then performed the split accordingly. We also removed the last row of the spectrogram, which is a choice that has a negligible impact since it represents only the highest 31.25 Hz band of the signal, but this allows us to have a power-of-2 input size which makes the design of G and D easier. Before the data are fed to our system, they are also normalized to be in the range $[-1,1]$.

We trained the GANs using stochastic gradient descent (SGD) and adopting the Adam optimizer, for 10 epochs with a batch size of 1 according to \cite{isola2016image}, updating G twice per each iteration to avoid a fast convergence of D \cite{lin2016pix2pix}. The networks' weights have been initialized from a normal distribution with zero mean and a standard deviation of 0.02 \cite{isola2016image}. The L1 loss has been added to the GAN loss using a scaling factor of 100 \cite{isola2016image}.

To enhance a speech signal with Pix2Pix, we first compute the T-F representation of it, and then we forward propagate the spectrogram magnitude through G. Finally, we reconstruct the signal with the inverse STFT using the phase of the noisy input.

\section{Experiments}

\subsection{Evaluation metrics}
\noindent The performance of our system is evaluated in terms of PESQ \cite{rix2001perceptual} (in particular the wide-band extension \cite{itu2005Wideband}), STOI \cite{taal2011algorithm}, and EER of ASV. PESQ and STOI have been chosen as they are the most used estimators of speech quality and speech intelligibility, respectively. The implementations used in this paper are from \cite{loizou2013speech} for PESQ and from \cite{taal2011algorithm} for STOI.

Regarding the ASV evaluation, we use the classical Gaussian Mixture Model - Universal Background Model (GMM-UBM) framework  \cite{sarkar2016text}, which is suitable for short utterances as in this work. We first built a general  model, UBM, which is a GMM trained with an expectation-maximization algorithm using a large amount of speech data not belonging to the target speakers. Then, a target speaker model for each specific pass-phrase and each speaker was derived by maximum a posteriori adaptation of the UBM. The approach of adapting UBM is used in order to have a well-trained model for a speaker even when there is no much data available. At this point, for a test utterance we calculate the log likelihood ratio between the claimant speaker model and the UBM.  The features extracted from the speech data are 57-dimensional mel-frequency cepstral coefficients (MFCCs), and the GMM mixture number is 512.

\subsection{Baseline methods}

We compare the results of our approach with other two methods we consider as baselines: STSA-MMSE and an Ideal Ratio Mask (IRM) based DNN-SE algorithms.

STSA-MMSE is a statistical-based SE technique, where the a priori signal to noise ratio (SNR) is estimated with the Decision-Directed approach \cite{ephraim1984speech} and the noise Power Spectral Density (PSD) is estimated with the noise PSD tracker in \cite{hendriks2010mmse}. The noise PSD estimate is initialized with the first 1000 samples of each utterance, assumed to be a speech-free region.

For the DNN-SE algorithm, we use the same procedure and parameters of \cite{kolbaek2017speech}. The IRM is estimated by using a DNN with three hidden layers of 1024 units each, and an output layer with 64 units. The input of the DNN is a 1845-dimensional feature vector, which is a robust representation of a frame that combines MFCCs, amplitude modulation spectrogram, relative spectral transform - perceptual linear prediction (RASTA-PLP), and gammatone filter bank energies, with their delta and double delta for a context of 2 past and 2 future frames. The training label is represented by the IRM, which is computed as in \cite{wang2014training} from the T-F representation based on a gammatone filter bank with 64 filters linearly spaced on a Mel frequency scale and with a bandwidth equal to one equivalent rectangular bandwidth  \cite{wang2006computational}. The system is trained for 30 epochs with SGD, using the mean square error as error function and a batch size of 1024. In order to enhance a test signal, the DNN provides an estimation of the IRM which is applied to the T-F representation of the noisy signal. Finally, the time domain signal is synthesized.

\subsection{Datasets}
\noindent We use two corpora, TIMIT \cite{garofolo1993darpa} and RSR2015 \cite{larcher2014text}, as follows:
\begin{itemize}
	\item Set 1 (TIMIT) - 4380 utterances of male speakers are used for UBM training.
    \item Set 2 (RSR2015) - Text ID from 2 to 30 of sessions 1, 4, and 7 for 50 male speakers (from m051 to m100) are selected to train Pix2Pix and DNN-SE.
    \item Set 3 (RSR2015) - Text ID 1 of sessions 1, 4, and 7 for 49 male speakers (from m002 to m050) are used to train the speaker models.
    \item Set 4 (RSR2015) - Sessions 2, 3, 5, 6, 8, and 9 of the same text ID and speakers used for training the models, are selected for evaluation. 
\end{itemize}


The choice of RSR2015 as the main database for training and testing can be seen as a compromise, because we were interested in the evaluation of an ASV system, which provides another objective measure of the performance, and RSR2015 is widely used for this task.

We used 5 different noise types to simulate real-life conditions: Babble, obtained by adding 6 random speech samples from the Librispeech corpus \cite{panayotov2015librispeech}; white Gaussian noise generated in MATLAB; Cantine, recorded by the authors; Market and Airplane, collected by Fondazione Ugo Bordoni (FUB) and available on request from the OCTAVE project~\cite{fuboctave}. Noise data, which were added to the utterances in Set 2, 3, and 4 at different SNR values, used for training and testing are different.

\subsection{Setup}

\noindent Inspired by \cite{kolboek2016speech}, we investigate two different kinds of Pix2Pix-based SE front-ends: 5 noise specific front-ends (NS-Pix2Pix), each of them trained on only one type of noise, and 1 noise general front-end (NG-Pix2Pix), trained on all types of noise. The same has been done for the DNN-SE front-ends, obtaining 5 noise specific front-ends (NS-DNN) and 1 noise general front-end (NG-DNN). For training, we add noise to clean speech at two different SNRs, 10 and 20 dB. With higher SNR it should be easier to train a G able to capture the underlying structure of the noisy input and generate a clean spectrogram, but a test with lower SNRs for training is worth to explore in the future. For testing, results for 5 different SNR conditions are reported: 0, 5, 10, 15, and 20 dB, as is commonly done for ASV, but an interesting future work is to test on lower SNRs, particularly for intelligibility evaluation. In addition, to find the behavior of the front-ends on noise free conditions, ASV performance on enhanced clean speech data is also reported. 

In all the tests, the performance of the following front-ends are presented: No enhancement (when no SE algorithm is used on noisy data), STSA-MMSE, NS-DNN, NS-Pix2Pix, NG-DNN, and NG-Pix2Pix. In total, three different tests have been conducted:
\begin{itemize}
	\item Test 1 - In the first test, we compute PESQ and STOI for the different front-ends to estimate speech quality and intelligibility.
    \item Test 2 - In the second test, the ASV system is trained with enhanced clean speech (except for the No enhancement front-end where clean speech is used) and tested on the 5 types of noise.
    \item Test 3 - The last test is performed to evaluate the effects of a multi-condition training on ASV. For No enhancement, STSA-MMSE, NS-DNN, and NS-Pix2Pix the speaker models are built from enhanced clean speech and one kind of enhanced noisy speech, while for NG-DNN and NG-Pix2Pix all kinds of noise are used.
\end{itemize}

\section{Results and Discussion}
\noindent The results of Test 1 are shown in Table \ref{tab:pesqstoi}. It is observed that the average PESQ scores of NS-Pix2Pix and NG-Pix2Pix are always better than the other front-ends. The best performance improvement is achieved between 5 and 15 dB SNR regardless of the noise type. At 20 dB, our approach outperforms the others on Market and White noises, but for Airplane noise STSA-MMSE is the best one, while for Babble and Cantine noises the absence of enhancement is superior indicating that all the SE techniques introduce an amount of distortion surpassing the benefit of noise reduction. At 0 dB, NG-Pix2Pix generally outperforms the noise specific version with an exception (Market noise) and its scores are close to DNN-SE ones.

In terms of STOI, Pix2Pix front-ends perform similarly to STSA-MMSE.
However, DNN-SE front-ends are superior in almost all the conditions, even though Pix2Pix front-ends achieve the same or very close results in some situations, e.g. low SNRs for Cantine and Market noises. At 20 dB, we observe the same behavior as the PESQ scores, where the evaluation of not enhanced signals gives a better outcome.

\begin{table}[ht]
  \caption{PESQ and STOI performance for the 5 front-ends: No enhancement (a), STSA-MMSE (b), NS-DNN (c), NS-Pix2Pix (d), NG-DNN (e), NG-Pix2Pix (f).}
  \label{tab:pesqstoi}
  \centering
  \resizebox{0.45\textwidth}{!}{%
  \begin{tabular}{ l | c | c c c c c | c || c c c c c | c }
    \toprule
    & & & & \multicolumn{2}{c}{PESQ} &  \multicolumn{1}{c}{} & &  \multicolumn{6}{c}{STOI} \\
    \toprule
     & SNR		&	0	& 5	&	10	&	15	&	20	&	mean &	0	& 5	&	10	&	15	&	20	&	mean\\
    \midrule
    
    \multirow{ 6}{*}{\rotatebox[origin=c]{90}{Airplane}}
    & (a) 	& 1.34 & 1.63 & 2.02 & 2.47 & 3.00 & 2.09 & 0.64 & 0.74 & 0.82 & \bf{0.88} & \bf{0.93} & 0.80\\
    & (b)	& 1.54 & 1.79 & 2.17 & 2.72 & \bf{3.26} & 2.30 & 0.66 & 0.74 & 0.81 & 0.87 & 0.91 & 0.80\\
    & (c) 	 & 1.65 & 1.94 & 2.30 & 2.73 & 3.16 & 2.36 & \bf{0.69} & \bf{0.76} & \bf{0.83} & \bf{0.88} & 0.92 & \bf{0.82}\\
    & \bf{(d)} & 1.57 & 2.02 & \bf{2.51} & \bf{2.91} & 3.18 & 2.44 & 0.66 & 0.75 & 0.81 & 0.85 & 0.89 & 0.79\\
    & (e) 	& 1.65 & 1.94 & 2.29 & 2.70 & 3.14 & 2.35 & \bf{0.69} & \bf{0.76} & 0.82 & 0.87 & 0.91 & 0.81\\
    & \bf{(f)} & \bf{1.67} & \bf{2.07} & \bf{2.51} & 2.88 & 3.13 & \bf{2.45} &  0.67 & 0.74 & 0.79 & 0.83 & 0.86 & 0.78\\

    \midrule
    \multirow{ 6}{*}{\rotatebox[origin=c]{90}{Babble}} 
    & (a) 	& 1.20 & 1.42 & 1.79 & 2.40 & \bf{3.13} & 1.99 & 0.44 & 0.56 & 0.67 & 0.77 & 0.85 & 0.66 \\
    & (b)	& 1.14 & 1.31 & 1.61 & 2.07 & 2.65 & 1.76 & 0.43 & 0.56 & 0.66 & 0.74 & 0.81 & 0.64 \\
    & (c)	& \bf{1.25} & 1.51 & 1.87 & 2.31 & 2.78 & 1.95 & \bf{0.50} & \bf{0.63} & \bf{0.72} & \bf{0.79} & \bf{0.86} & \bf{0.70} \\
    & \bf{(d)} & 1.20 & 1.48 & 1.98 & 2.52 & 2.93 & 2.02 & 0.46 & 0.59 & 0.71 & 0.78 & 0.83 & 0.67 \\
    & (e) & 1.24 & \bf{1.52} & 1.88 & 2.31 & 2.78 & 1.95 & 0.49 & 0.62 & \bf{0.72} & \bf{0.79} & 0.85 & \bf{0.70} \\
    & \bf{(f)} & 1.20 & 1.49 & \bf{2.00} & \bf{2.53} & 2.93 & \bf{2.03} & 0.46 & 0.60 & 0.71 & 0.77 & 0.82 & 0.67 \\
    
    \midrule
    \multirow{ 6}{*}{\rotatebox[origin=c]{90}{Cantine}} 
    & (a)	& 1.35 & 1.65 & 2.07 & 2.57 & \bf{3.30} & 2.19 & 0.54 & 0.66 & 0.75 & \bf{0.83} & \bf{0.90} & 0.74 \\
    & (b)	& 1.38 & 1.68 & 2.12 & 2.67 & 3.23 & 2.22 & 0.55 & 0.66 & 0.74 & 0.82 & 0.87 & 0.73 \\
    & (c) 	& 1.46 & 1.75 & 2.15 & 2.63 & 3.12 & 2.22 & 0.59 & \bf{0.69} & 0.76 & \bf{0.83} & 0.89 & 0.75 \\
    & \bf{(d)} & 1.45 & 1.84 & 2.38 & \bf{2.82} & 3.13 & 2.32 & 0.58 & 0.68 & 0.75 & 0.80 & 0.85 & 0.73 \\
    & (e) & 1.47 & 1.77 & 2.18 & 2.64 & 3.11 & 2.24 & \bf{0.60} & \bf{0.69} & \bf{0.77} & \bf{0.83} & 0.89 & \bf{0.76} \\
    & \bf{(f)} & \bf{1.49} & \bf{1.91} & \bf{2.43} & 2.81 & 3.08 & \bf{2.34} & 0.59 & \bf{0.69} & 0.75 & 0.80 & 0.84 & 0.73 \\
    
    \midrule
    \multirow{ 6}{*}{\rotatebox[origin=c]{90}{Market}}
    & (a) 	& 1.26 & 1.51 & 1.89 & 2.38 & 3.04 & 2.02 & 0.51 & 0.62 & 0.73 & 0.81 & \bf{0.88} & 0.71 \\
    & (b)	& 1.24 & 1.45 & 1.76 & 2.22 & 2.79 & 1.89 & 0.51 & 0.62 & 0.71 & 0.79 & 0.85 & 0.70 \\
    & (c)	& 1.35 & 1.63 & 2.00 & 2.46 & 2.94 & 2.08 & \bf{0.56} & \bf{0.67} & \bf{0.75} & \bf{0.82} & \bf{0.88} & \bf{0.73} \\
    & \bf{(d)} & \bf{1.36} & 1.71 & 2.21 & \bf{2.72} & \bf{3.09} & \bf{2.22} & 0.55 & 0.66 & 0.74 & 0.80 & 0.85 & 0.72 \\
    & (e) & \bf{1.36} & 1.63 & 2.00 & 2.45 & 2.93 & 2.07 & \bf{0.56} & \bf{0.67} & \bf{0.75} & \bf{0.82} & \bf{0.88} & \bf{0.73} \\
    & \bf{(f)} & 1.35 & \bf{1.72} & \bf{2.24} & 2.68 & 3.02 & 2.20 & \bf{0.56} & \bf{0.67} & 0.74 & 0.79 & 0.83 & 0.72 \\
    
    \midrule
    \multirow{ 6}{*}{\rotatebox[origin=c]{90}{White}} 
    & (a) 	& 1.15 & 1.31 & 1.60 & 2.01 & 2.57 & 1.73 & 0.50 & 0.61 & 0.72 & 0.81 & \bf{0.89} & 0.71 \\
    & (b)	& 1.35 & 1.58 & 1.88 & 2.25 & 2.71 & 1.95 & 0.53 & 0.63 & 0.73 & 0.81 & 0.87 & 0.72 \\
    & (c)	& \bf{1.38} & 1.66 & 2.00 & 2.39 & 2.88 & 2.06 & \bf{0.58} & \bf{0.67} & \bf{0.75} & \bf{0.82} & 0.88 & \bf{0.74} \\
    & \bf{(d)} & 1.23 & 1.54 & 2.11 & \bf{2.74} & \bf{3.14} & 2.15 & 0.53 & 0.64 & 0.73 & 0.80 & 0.86 & 0.71 \\
    & (e) & 1.35 & 1.63 & 1.96 & 2.29 & 2.65 & 1.98 & 0.57 & 0.66 & 0.74 & 0.81 & 0.88 & 0.73 \\
    & \bf{(f)} & 1.32 & \bf{1.69} & \bf{2.22} & 2.68 & 3.01 & \bf{2.19} & 0.55 & 0.65 & 0.73 & 0.78 & 0.83 & 0.71 \\
    
    \bottomrule
  \end{tabular}}
  
\end{table}

 \begin{figure*}[t]
   \centering
   \includegraphics[width=\linewidth, height=2.62cm]{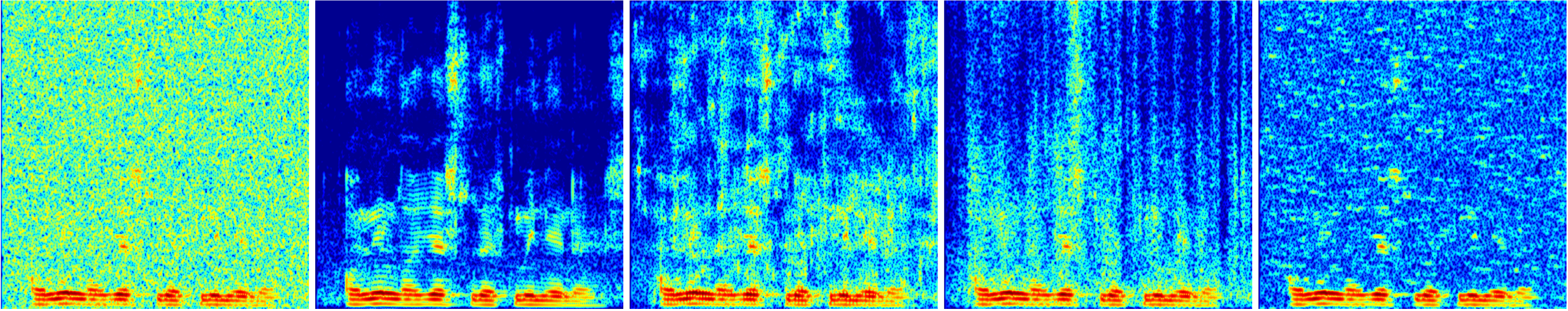}
   \caption{From left to right: noisy spectrogram (White noise at 0 dB SNR); clean spectrogram; spectrogram of the signal enhanced with NG-Pix2Pix; spectrogram of the signal enhanced with NG-DNN; spectrogram of the signal enhanced with STSA-MMSE.}
   \label{fig:spectrograms}
 \end{figure*}

\begin{table}[th]
  \caption{ASV performance in terms of EER on clean speaker model}
  \label{tab:sv-clean}
  \centering
  \resizebox{0.45\textwidth}{!}{%
  \begin{tabular}{ l | r| c c c c c c | c }
    \toprule
     & SNR		&	0	& 5	&	10	&	15	&	20 & clean	&	mean \\
    \midrule

\multirow{ 6}{*}{\rotatebox[origin=c]{90}{Airplane}}
& No enhancement 	& 21.09 & 15.99 & 13.61 & 11.66 & 9.18 & 6.99 & 13.08\\ 
												
& STSA-MMSE	& 17.69 & 12.58 & 8.17 & 6.53 & 6.27 & 5.80 & 9.51\\
												
& NS-DNN 	& 16.99 & 10.55 & 7.48 & 6.99 & 6.15 & 6.12 & 9.05 \\
												
& \bf{NS-Pix2Pix} & 17.19 & 8.84 & \bf{5.44} & 5.05 & \bf{4.63} & \bf{3.74}  & 7.48\\
												
& NG-DNN 	& 15.99 & 8.99 & 6.12 & 6.12 & 5.58 & 5.67 & 8.08\\
												
& \bf{NG-Pix2Pix} & \bf{15.31} & \bf{7.89} & 5.58 & \bf{4.77} & 4.76 & 5.44 & \bf{7.29}\\
												
\midrule
\multirow{ 6}{*}{\rotatebox[origin=c]{90}{Babble}}
& No enhancement & 19.05 & 14.63 & 11.69 & 11.04 & 9.18 & 6.99 & 12.10 \\
												
& STSA-MMSE	& 29.04 & 20.40 & 12.59 & 7.82 & 6.29 & 5.80 & 13.66\\
												
& NS-DNN 	& 17.01 & 10.54 & 7.82 & 6.46 & 6.12 & 5.78 & 8.96\\
												
& \bf{NS-Pix2Pix} &  18.83 & 11.22 & 7.62 & \bf{5.70} & 5.10 & \bf{4.08} & 8.76 \\
												
& NG-DNN & \bf{16.67} & \bf{10.39} & \bf{7.50} & 6.34 & 5.78 & 5.67 & \bf{8.73}\\
												
& \bf{NG-Pix2Pix} & 21.05 & 13.64 & 8.50 & 5.97 & \bf{4.76} & 5.44 & 9.90\\ 
												
\midrule
\multirow{ 6}{*}{\rotatebox[origin=c]{90}{Cantine}}
& No enhancement 	& 20.72 & 19.20 & 14.74 & 11.81 & 8.50 & 6.99 & 13.66\\
												
& STSA-MMSE	& 19.09 & 12.37 & 8.16 & 6.80 & 6.12 & 5.80 & 9.72\\
												
& NS-DNN 	& 18.71 & \bf{8.58} & 6.12 & 5.49 & 5.31 & 5.10 & 8.22\\
												
& \bf{NS-Pix2Pix} &  \bf{17.33} & 9.18 & \bf{5.44} & \bf{5.10} & 5.10 & \bf{4.16} & \bf{7.72}\\
												
& NG-DNN & 19.94 & 9.18 & 6.12 & 5.78 & 5.44 & 5.67 & 8.69\\
												
& \bf{NG-Pix2Pix} & 17.57 & 8.84 & 5.73 & 5.31 & \bf{4.76} & 5.44 & 7.94\\
												
\midrule
\multirow{ 6}{*}{\rotatebox[origin=c]{90}{Market}}
& No enhancement 	& 29.40 & 20.07 & 15.00 & 11.96 & 8.93 & 6.99 & 15.39\\
												
& STSA-MMSE	& 25.51 & 17.35 & 11.90 & 8.28 & 7.35 & 5.80 & 12.70\\
												
& NS-DNN 	& 21.43 & \bf{9.86} & \bf{6.88} & 6.46 & 5.78 & 5.92 & 9.39\\
												
& \bf{NS-Pix2Pix} &  \bf{17.91} & 10.33 & 7.14 & \bf{5.92} & 5.17 & \bf{3.61} & \bf{8.35}\\
												
& NG-DNN & 21.77 & 10.59 & 7.48 & 6.22 & 5.76 & 5.67 & 9.58\\
												
& \bf{NG-Pix2Pix} & 19.58 & 11.22 & 7.48 & 6.12 & \bf{5.07} & 5.44 & 9.15\\
												
\midrule
\multirow{ 6}{*}{\rotatebox[origin=c]{90}{White}} 
& No enhancement 	& 45.90 & 43.20 & 34.61 & 26.28 & 16.91 & 6.99 & 28.98 \\
												
& STSA-MMSE	& 30.95 & 21.17 & 13.95 & 10.20 & 8.50 & 5.80 & 15.10\\ 
												
& NS-DNN 	& 39.46 & 20.75 & 9.86 & 7.82 & 6.12 & 6.02 & 15.01\\
												
& \bf{NS-Pix2Pix} &  40.48 & 28.23 & 12.45 & 7.86 & 6.46 & 6.46 & 16.99\\
												
& NG-DNN & 40.14 & 21.77 & 10.88 & 8.16 & 6.80 & 5.67 & 15.57\\
												
& \bf{NG-Pix2Pix} & \bf{30.61} & \bf{17.33} & \bf{9.40} & \bf{7.14} & \bf{5.78} & \bf{5.44} & \bf{12.62}\\

    \bottomrule
  \end{tabular}}
  
\end{table}

\begin{table}[ht]
  \caption{ASV performance in terms of EER on multi-condition speaker model}
  \label{tab:sv-mc}
  \centering
  \resizebox{0.45\textwidth}{!}{%
  \begin{tabular}{ l | r| c c c c c c | c }
    \toprule
     & SNR		&	0	& 5	&	10	&	15	&	20 & clean	&	mean \\
    \midrule
    
    \multirow{ 6}{*}{\rotatebox[origin=c]{90}{Airplane}}
    & No enhancement 	& 32.28 & 26.87 & 21.10 & 16.38 & 9.86 & 5.83 & 18.72 \\
    & STSA-MMSE 	& 25.51 & 15.48 & 8.16 & 6.12 & 5.44 & 5.44 & 11.03 \\
    & NS-DNN 		& 14.78 & 8.26 & 5.44 & 5.53 & 4.76 & 4.76 & 7.26 \\
    & \bf{NS-Pix2Pix} 	& 16.67 & 7.14 & 5.10 & \bf{4.03} & \bf{3.78} & 4.42 & 6.86 \\
    & NG-DNN 		& \bf{11.38} & \bf{6.12} & \bf{4.78} & 4.72 & 4.23 & \bf{4.00} & \bf{5.87} \\
    & \bf{NG-Pix2Pix} 	& 13.27 & 6.43 & 5.78 & 5.44 & 5.27 & 4.78 & 6.83\\

    \midrule
    \multirow{ 6}{*}{\rotatebox[origin=c]{90}{Babble}}
    & No enhancement 	& 21.77 & 15.37 & 11.93 & 9.52 & 8.16 & 6.12 & 12.15 \\
    & STSA-MMSE		& 33.50 & 23.13 & 16.23 & 12.63 & 8.84 & 7.12 & 16.91 \\
    & NS-DNN		& 16.26 & 9.52 & 6.99 & 6.08 & 5.78 & 5.17 & 8.30 \\
    & \bf{NS-Pix2Pix}	& 20.75 & 10.88 & 6.12 & \bf{4.76} & \bf{4.08} & 4.36 & 8.49 \\
    & NG-DNN		& \bf{16.00} & \bf{9.18} & \bf{5.44} & \bf{4.76} & \bf{4.08} & \bf{4.00} & \bf{7.19} \\
    & \bf{NG-Pix2Pix}	& 21.72 & 12.44 & 6.46 & 5.34 & 5.22 & 4.78 & 9.33 \\
    
    \midrule
    \multirow{ 6}{*}{\rotatebox[origin=c]{90}{Cantine}}
    & No enhancement	& 24.11 & 17.22 & 12.93 & 10.88 & 9.18 & 7.48 & 13.63 \\
    & STSA-MMSE		& 19.05 & 12.59 & 8.21 & 6.91 & 6.12 & 6.32 & 9.87 \\
    & NS-DNN		& 12.93 & 5.91  & \bf{4.42} & 4.25 & 4.27 & \bf{3.78} & 5.93 \\
    & \bf{NS-Pix2Pix}	& 14.29 & 6.87 & 4.76 & \bf{4.00} & \bf{4.08} & 4.76 & 6.46 \\
    & NG-DNN		& \bf{11.61} & \bf{5.78} & 5.10 & 4.57 & \bf{4.08} & 4.00 & \bf{5.86} \\
    & \bf{NG-Pix2Pix}	& 14.10 & 7.48 & 5.44 & 5.44 & 5.27 & 4.78 & 7.08 \\
    
    \midrule
    \multirow{ 6}{*}{\rotatebox[origin=c]{90}{Market}}
    & No enhancement 	& 36.05 & 26.06 & 18.37 & 13.32 & 9.18 & 5.44 & 18.07 \\
    & STSA-MMSE		& 29.25 & 21.07 & 13.95 & 10.98 & 7.82 & 6.67 & 14.97 \\
    & NS-DNN		& 19.33 & \bf{8.16} & 6.24 & 5.41 & 4.53 & 4.29 & 7.99 \\
    & \bf{NS-Pix2Pix}	& 18.49 & 9.18 & 5.82 & \bf{4.42} & \bf{3.74} & 4.76 & 7.74 \\
    & NG-DNN		& \bf{18.37} & \bf{8.16} & \bf{5.78} & 4.44 & 4.42 & \bf{4.00} & \bf{7.53} \\
    & \bf{NG-Pix2Pix}	& 19.30 & 9.37 & 6.37 & 5.44 & 5.10 & 4.78 & 8.39\\
    
    \midrule
    \multirow{ 6}{*}{\rotatebox[origin=c]{90}{White}} 
    & No enhancement	& 35.88 & 24.40 & 18.37 & 15.81 & 14.97 & 5.85 & 19.21 \\
    & STSA-MMSE		& 30.95 & 20.07 & 7.48 & 6.46 & 6.46 & 4.76 & 12.70 \\
    & NS-DNN		& 27.21 & \bf{9.52} & \bf{6.12} & \bf{5.02} & 4.65 & 5.78 & 9.72 \\
    & \bf{NS-Pix2Pix}	& 39.37 & 23.81 & 10.20 & 6.46 & 5.95 & 6.44 & 15.37 \\
    & NG-DNN		& \bf{26.19} & 11.22 & 7.14 & 5.10 & \bf{4.08} & \bf{4.00} & \bf{9.62} \\
    & \bf{NG-Pix2Pix}	& 30.41 & 14.29 & 8.84 & 6.60 & 5.78 & 4.78 & 11.78 \\
    
    \bottomrule
  \end{tabular}}
  
\end{table}

The ASV performances (Tests 2 and 3) are reported in Tables \ref{tab:sv-clean} and \ref{tab:sv-mc}, where the results of the baseline systems are from \cite{Yu2017Adversarial}. For the clean speaker models, Pix2Pix front-ends generally outperform the baseline methods. One exception is seen for Babble noise, where the NG-DNN front-end gives an EER of 8.73\%, marginally better than NS-Pix2Pix (8.76\%). At low SNR, DNN-SE front-ends sometimes show better results than Pix2Pix, but overall our approach can be considered superior.

On the other hand, the performances of DNN-SE front-ends on multi-condition training are generally better, which presents a substantial improvement if compared with the clean speaker model situation. Our approach is generally better than STSA-MMSE, although the NS-Pix2Pix front-end shows lower performance when it deals with white noise.

In general, Pix2Pix can be considered competitive with DNN-SE (better PESQ and EER on the clean speaker models, but worse STOI and EER for multi-condition training) and overall superior to STSA-MMSE.

Figure \ref{fig:spectrograms} shows the spectrograms of a noisy utterance (White noise at 0 dB SNR), together with its clean and enhanced versions with NG-Pix2Pix, NG-DNN, and STSA-MMSE. It is observed that the spectrogram enhanced by the cGAN approach preserves the structure of the original signal better than the other SE techniques, while at the same time more noises remain especially at high frequency regions, as compared with NG-DNN. The spectrogram enhanced by STSA-MMSE is snowy all over the places.

\section{Conclusion}
\noindent In this paper we investigated the use of conditional generative adversarial networks (cGANs) for speech enhancement. We adapted the Pix2Pix framework, intended to solve generic image-to-image translation problems, and evaluated the performance of this approach in terms of estimated speech perceptual quality and speech intelligibility, together with equal error rate of a Gaussian Mixture Model - Universal Background Model based speaker verification system.  The results we obtained allow us to conclude that cGANs are a promising technique for speech denoising, being globally superior to the classical STSA-MMSE algorithm, and comparable to a DNN-SE algorithm. 

Future work includes a more extensive evaluation of the framework in more critical SNR situations, and modifications aiming at making it specific for the task. For example, a model with G generating a small size output window from a fixed number of successive frames can be built as it is often done in deep neural networks for speech processing, and a specific perceptual loss to be added to the cGAN loss can be designed.

\section{Acknowledgements}

\noindent The authors would like to thank Hong Yu for providing data and speaker verification results for the baseline systems and Morten Kolb{\ae}k  for his assistance and software used for the speaker verification and DNN speech enhancement baseline systems.

This work is partly supported by the  Horizon 2020 OCTAVE Project (\#647850), funded by the Research European Agency (REA) of the European Commission, and the iSocioBot project, funded by the Danish Council for Independent Research - Technology and Production Sciences (\#1335-00162).



\bibliographystyle{IEEEtran}

\newpage
\bibliography{gan_daniel}


\end{document}